\documentclass[twocolumn,showpacs,secnumarabic,amssymb,aps,prl,superscriptaddress]{revtex4}

\usepackage{graphicx}
\usepackage{enumerate}

\begin{document}
\title{Fault-Tolerant Topological One-Way Quantum Computation with Probabilistic Two-Qubit Gates}%

\author{Keisuke Fujii}
\affiliation{ Department of Nuclear Engineering, Kyoto University, Kyoto 606-8501, Japan}
\author{Yuuki Tokunaga}
\affiliation{ NTT Information Sharing Platform Laboratories, NTT Corporation, 3-9-11 Midori-cho, Musashino, Tokyo 180-8585, Japan}
\date{\today}%

\begin{abstract}
We propose a scalable way to construct 
a 3D cluster state for fault-tolerant topological one-way computation
 (TOWC) even if the entangling two-qubit gates
succeed with a small probability.
It is shown that fault-tolerant TOWC can be performed 
with the success probability of the two-qubit gate such as 0.5 (0.1)
provided that the unheralded error probability of the two-qubit gate
is less than $0.040\%$ ($0.016\%$). 
Furthermore, the resource usage is considerably 
suppressed compared to the conventional fault-tolerant schemes with
probabilistic two-qubit gates.
\end{abstract}

\pacs{03.67.Pp,03.67.Lx}
\maketitle

{\it Introduction.---}
Quantum computation has a great deal of potential,
enabling us to solve some sorts of problems,
which are thought to be intractable with classical computers.
However, quantum states are very fragile 
against environmental interaction
and suffer from decoherence.
Fortunately, the decoherence effect can be counteracted 
by fault-tolerant quantum computation
\cite{Shor95,DiVincenzo-Shor96,Gottesman97,Kitaev97a,Preskill98,Knill98,Aharanov-Ben-Or99,Knill96a,Steane03}.
Actually, if the amount of noise per gate is smaller than
a certain value, so called noise threshold, 
quantum computation can be realized 
to arbitrary accuracy with only a polynomial overhead
\cite{Kitaev97a,Knill98,Preskill98,Aharanov-Ben-Or99}. 
The noise thresholds have been calculated to be about $10^{-4}-10^{-2}$
for several fault-tolerant schemes
under varying degrees of assumption and rigor 
\cite{Steane03,Knill05,Aliferis,Raussendorf07,FY10}.

In fault-tolerant theory,
it is often assumed that
two-qubit gates are deterministic and can be performed between spatially separate qubits.
Under these assumptions,
the noise thresholds have been calculated to be $1\%-3\%$ numerically \cite{Knill05,FY10}
and $\sim 10^{-3}$ rigorously \cite{Aliferis}.
On the other hand, in a wide range of physical systems
(e.g., ion trap, quantum dots, Josephson-junction qubits, 
and neutral atoms in optical lattices),
interaction for gate operations is restricted to the nearest neighbors.
This requirement would be fulfilled
naturally in one-way computation (OWC),
where computational resource states, so called cluster states, are
constructed with only the neighbor two-qubit gates \cite{Raussendorf01}.
It has been found that
fault-tolerant quantum computation
can be performed on a three-dimensional (3D) cluster state via OWC \cite{Raussendorf07}.
Another important physical property 
is nondeterminism in the two-qubit gates.
Especially in linear optics,
two-qubit gates are intrinsically non-deterministic
due to the linearity of the interaction \cite{KLM01,Pittman01,Browne05}.
In other systems,
it is also often the case that 
a large amount of errors
can be detected in heralded ways,
and one can postselect the successful events \cite{Barrett05,Duan05B}.
In Ref. \cite{Barrett05}, the unheralded error probability of the postselected
gates is estimated to be about $10^{-4}$, where the success probability 
of the gates becomes 0.1--0.2.

We explore fault-tolerant quantum computation
with such probabilistic two-qubit gates (PTQGs).
The noisy PTQGs are 
characterized by two types of errors,
namely heralded and unheralded errors.
The PTQG succeeds with the probability $p_{s}$
(i.e., heralded errors are detected with the probability $1-p_{s}$),
but the unheralded errors still remain with probability $p_{u}$.
The first approach in this line
is proposed based on 
the unique future of OWC,
which promises fault-tolerance
under $(p_{\rm loss},p_{u})=(10^{-3},10^{-4})$ by using 
linear optical fusion gates with success probability $p_{s} =1/2$ 
\cite{Nielsen04,Dawson06,Cho07}.
The recent study has revealed that
$(p_{s}, p_{u})=(0.9, 0.03\%)$ or $(0.95, 0.3\%)$ are sufficient 
for fault-tolerant computation \cite{Goto09},
based on Knill's error-correcting $C_{4}/C_{6}$ architecture \cite{Knill05}.
However, it has not been clear whether
fault-tolerant computation can be implemented 
with PTQGs under $p_s < 1/2$ or not.
 
In this Letter, we construct a fault-tolerant scheme with 
PTQGs based on
topological one-way computation (TOWC)
on the 3D cluster state,
which works well even with a small success probability $p_{s} < 1/2$. 
The present scheme can be obtained by combining
novel techniques which have been developed 
in the field of OWC to date.
The key features are as follows:
(i) Near-deterministic cluster-state construction
based on the so called divide and conquer approach \cite{Dawson06,Browne05,Duan05}.
(ii) Fault-tolerant TOWC
on the 3D cluster state \cite{Raussendorf07}.
(iii) Counterfactual error correction
with the indirect measurements \cite{Varnava06}
to suppress the error accumulation.
A specific shape of cluster states,
say star-cluster, 
are prepared with postselection in parallel.
Then, they are connected by using 
PTQGs near-deterministically
to form the 3D cluster state 
for topological computation.
The error accumulation during the near-deterministic connection
is made small
for a reasonably small success probability $p_{s} \sim 0.1$.
Then, fault-tolerant computation is 
performed by single qubit measurements
on the 3D cluster state.
It is shown that
fault-tolerant TOWC is possible with, e.g.,
$(p_{s}, p_{u})= ( 0.9 , 6.0 \times 10^{-4})$, 
$(0.5, 4.0 \times  10^{-4})$, or $(0.1, 1.6 \times 10^{-4})$.
Furthermore, the total overhead is significantly reduced
compared to them.

\begin{figure}
\centering
\includegraphics[width=85mm]{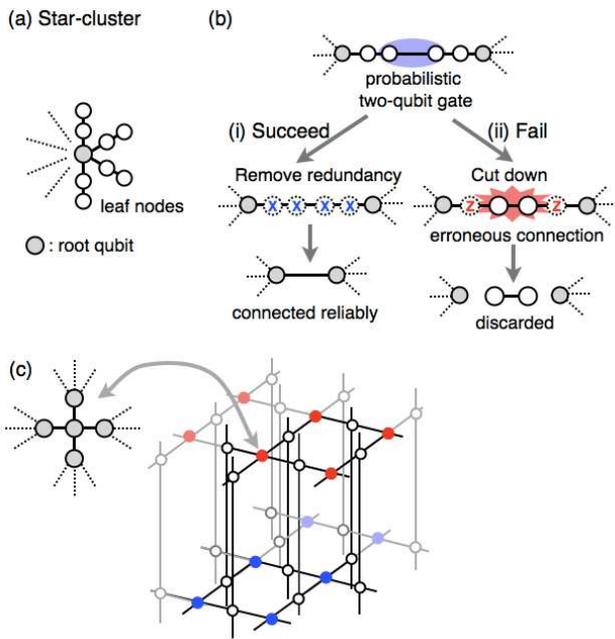}
\caption{(a) Star cluster. (b) Near-deterministic connection. (c) Four neighboring root nodes are connected so as to form the 3D cluster state for TOWC.}
\label{StarCluster}
\end{figure}

{\it Protocol 1 (P{\rm 1}).---}
We first prepare the star cluster, 
which consists of one ``root qubit" located at the center
and the surrounding $L$ ``leaf nodes" as depicted in Fig. \ref{StarCluster} (a).
Two neighboring root qubits are connected 
by using the redundant leaf qubits
via the PTQG.
We basically follow the method in Refs. \cite{Nielsen04,Dawson06}, 
so called divide and conquer approach.
In the case of the linear optical fusion gate,
the failure event results in a disconnected cluster state \cite{comment}.
Here, we consider a more general PTQG:
even if the PTQG fails,
the cluster state may be still connected erroneously
and hence have to be disconnected.
To resolve this,
if the two qubit gate fails,
the connection is discarded by measuring 
the adjacent qubits in $Z$ basis as seen in Fig. \ref{StarCluster} (b) (ii).
After several trials,
one can reliably connect a leaf node between 
the neighboring root qubits.
To connect two root qubits directly,
the redundant qubits between them
are measured in the $X$ basis as shown in Fig. \ref{StarCluster} (b) (i).
By repeating these procedures
one can connect the root qubit with the four neighboring root qubits, 
forming the 3D cluster state for TOWC,
as depicted in Fig. \ref{StarCluster} (c).
The redundant leaf nodes after the successful connections of four leaf nodes
are cut down by the $Z$ basis measurements 
similarly to the unsuccessful case.
The number of the leaf nodes $L$ has to be sufficiently large
so that the probability $p_{f}$ to fail the connections of 
four root qubits is sufficiently small.
The failure probability
can be given by
\begin{eqnarray}
p_{f} =  \sum _{k=0}^{3} 
\left( \begin{array}{c}
L \\ k
\end{array} \right)
p_{s}^{k} (1-p_{s})^{L-k} ,
\end{eqnarray}
where $k$ corresponds to the number of successful connections.
The erroneous connections with the probability $p_{f}$ 
are used as though they had succeeded.
Such rare events can be treated as detected computational errors,
which are corrected during TOWC like qubit-loss errors \cite{Barrett10}.
Specifically, in the case of $p_{s} =0.9,0.5,0.1$,
$L = 7 ,17, 97$ are sufficient for $p_{f} < 1\%$, respectively.
As shown in Ref. \cite{Barrett10},
detected errors of $\sim 1\%$ only slightly change
the noise threshold.
Otherwise,
one can also treat the failures straightforwardly as undetected computational errors, 
while it requires $p_{f} \sim p_{u}$ and results in a slight
deterioration of the performance.
After the near-deterministic construction of the 3D cluster,
TOWC can be performed by measuring the root qubits.
The errors introduced during the near-deterministic connection
appear in the measurement outcomes of the root qubits, 
which renormalize the measurement error probability of the root qubits. 

To make a clear exposition of the error accumulation process,
we first assume the unheralded error
affects on the measurement outcomes 
with probability $p_{u}$ independently for each qubit.
We will properly treat the preparation, measurement,
and two-qubit gate errors later on.  
Upon this independent noise model,
the renormalized error probability for the root qubit
with the four successful connections
can be calculated in the leading order as
\begin{eqnarray}
p_{r} = p_{u} + 4 \times 2p_{u} 
+ (L-4) p_{u} .
\label{eq-root-1}
\end{eqnarray}
The first term comes from the error 
on the root qubit itself.
The second and third terms are responsible for 
the error propagations from
$X$ (former) and $Z$ (latter) bases measurements 
for the successful and unsuccessful (or redundant) cases, respectively.
Then, if the renormalized error probability $p_{r}$  is
sufficiently smaller than
the threshold value of the surface code,
TOWC can be performed with arbitrary accuracy.
As seen from Eq. (\ref{eq-root-1}), 
$p_{r}$ depends on
$L \sim \mathcal{O} (1/p_{s})$,
which results in an extensive error accumulation
with a small success probability. 

{\it Protocol 2 (P{\rm 2}).---}
In the case of a low success probability,
the number of leaf nodes $L$ has to be 
large for the near-deterministic connection. 
In such a case, the $L$ dependence in Eq. (\ref{eq-root-1})
causes an extensive error accumulation.
We next develop a way to reduce the $L$
dependence by utilizing the counterfactual error correction scheme \cite{Varnava06}.

\begin{figure}
\centering
\includegraphics[width=85mm]{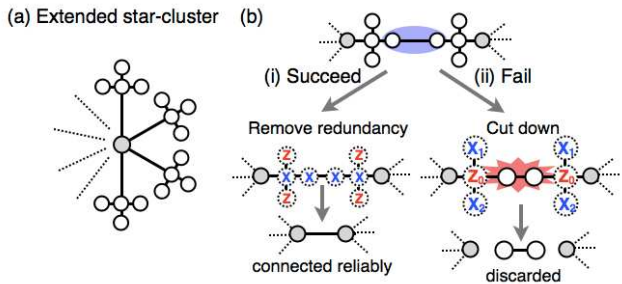}
\caption{(a) Extended star cluster. (b) Near-deterministic connection with
the counterfactual error correction by the indirect measurements.}
\label{ExStarCluster}
\end{figure}
The star cluster is now extended 
as shown in Fig. \ref{ExStarCluster} (a).
Similarly to the previous case,
the neighboring root qubits are connected
with the PTQGs via the leaf nodes.
If the connection succeeds,
the redundant qubits between the two root qubits
are measured $X$ and $Z$ bases 
as seen in Fig. \ref{ExStarCluster} (b) (i).
On the other hand,
if the PTQG fails,
the neighboring qubits are measured 
in $Z$ basis to discard the erroneous connection.
In addition,
the two qubits connected to the $Z$-measured qubit 
are measured in the $X$ basis 
as shown in Fig. \ref{ExStarCluster} (b) (ii).
Since these three qubits
are stabilized by 
the following operators
$\{ S_{1}= Z_{0}X_{1},S_{2}=Z_{0}X_{2} \}$,
one can correct an error on the $Z_{0}$ measurement outcome
by using the outcomes of the $X_{1,2}$ measurements,
which are the indirect measurements of the observable $Z_{0}$ \cite{Varnava06}.
This can be also understood from the fact that
$\{ S_{1}, S_{2} \}$ forms a three qubit repetition code,
where $X$ error on the qubit 0
and $Z$ errors on the qubit 1 and 2 can be corrected.
Then,
the renormalized error probability
of the root qubit can be rewritten as
\begin{eqnarray}
p_{r} = p_{u} +  4 \times 4 p_{u} 
+ (L-4) \times  3p_{u}^2 .
\label{eq-root-2}
\end{eqnarray}
If $p_{u}$ is sufficiently small, more precisely $p_{u} \ll 1/L$,
one can neglect the $L$ dependence in $p_{r}$.
The trade-off is that
the additional errors are added 
in the successful case
as seen in the second term
of Eq. (\ref{eq-root-2}).
By comparing Eqs. (\ref{eq-root-1}) and (\ref{eq-root-2}),
one can understand that
the counterfactual correction draws a certain figure of merit for $L>12$
in this model.

\begin{figure}
\centering
\includegraphics[width=65mm]{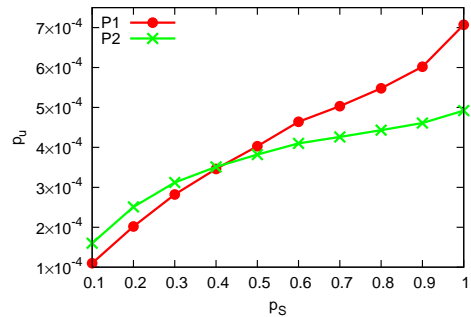}
\caption{The unheralded error probability $p_{u}$ which satisfies $p_{r} = 2\%$
is plotted as a function of $p_s$ for each {\it P}1 (circle) and {\it P}2 (cross).}
\label{Threshold}
\end{figure}

{\it Noise threshold and resource overhead.---}
So far, we have considered
the simple independent noise model.
We further proceed to the detailed calculation
of the renormalized error probability 
incorporating the preparation, measurement,
and unheralded two-qubit gate errors.
Here we adopt the following noise model:
(i) The noisy PTQG
under the condition of success is modeled by
an ideal one followed by $A \otimes B$ errors
with probabilities $p_{u}/15$ ($ A, B = I, X, Y, Z $, and $ A \otimes B
\not= I \otimes I $).
(ii) The erroneous preparation and measurement of physical qubits are 
executed by ideal operations followed by depolarization
with the error probabilities $p_{P}$ and $p_{M}$, respectively.
We can calculate the contributions from
the preparation and measurement errors 
similarly to the independent noise.
This is because the preparation errors 
are commutable with the CZ gates
for constructing the cluster states, 
and the measurement errors are the noise that we have just assumed in
the independent noise model.
On the other hand,
the two-qubit gate errors
behave somewhat complicatedly,
since we have to calculate all the commutation relations
of each error operator with the posterior two-qubit gates.
Moreover, 
whether a leaf node is successfully connected or not
is determined probabilistically.
Therefore, in order to evaluate 
the two-qubit errors,
we resort to the numerical calculations as done also in Ref. \cite{Dawson06}.
Then, the renormalized error probability of the root qubit
for {\it P}1 is given by
\begin{eqnarray}
p_{r} ^{(1)}
&=&  (7.7+ 0.64L)p_{u} + 
p_{P}+p_{M} \nonumber \\
&&
 + 4 \times 2(p_{M} +p_{P})
+ (L-4) p_{M},
\end{eqnarray}
where $p_{r}$ is responsible for the independent errors.
If there are only independent errors,
the threshold condition is given by $p_{r} = 3.3\%$,
which is the threshold value of the surface code \cite{Wang03Ohno04}.
However, there are also correlated errors,
which act on different root qubits simultaneously.
Fortunately, 
a large part of the correlated errors (i.e. on the nearest neighbor sites)
play no role and can be treated as independent errors 
because they are corrected independently in TOWC \cite{Raussendorf07}.
Then, as a result of numerical calculation, 
the probability of the truly correlated errors
(i.e. on the second-nearest neighbor sites) is 
given by at least one order of magnitude smaller
than the independent one. 
In such a situation, 
the correlated errors 
only slightly change the threshold.
In fact, $p_{r} = 2.05\%$ gives the threshold
when the correlated error probability is $0.26\%$ \cite{Raussendorf07}.
Thus the threshold condition is fairly given by $p_{r} < 2\%$,
for such weakly correlated noise.
In Fig. \ref{Threshold}, we plot the unheralded error probability $p_{u}$,
which satisfies $p_{r} = 2.0\%$,
as a function of the success probability $p_s$,
where $p_{P} = p_{M} = p_{u}$ is adopted for concreteness.
The region $(p_{s},p_{u})$ below the threshold curve
guarantees fault-tolerant quantum computation.
The exact noise threshold of $p_{u}$, of course, should be determined 
via the optimization process
by performing the detailed numerical simulation including 
all detected, independent, and correlated errors.
As for {\it P}2,
the renormalized error probabilities can be calculated similarly as 
\begin{eqnarray}
p_{r} ^{(2)}
&=&  (11+ 0.90L) p_{u} + 
p_{P}+p_{M} 
 + 4 \times 2(2p_{M} +p_{P})
 \nonumber \\ &&
+(L-4)(3p_{M}+p_{P})(p_{M}+p_{P}).
\end{eqnarray}
Here, $p_{r}^{(2)}$ is less depending on the size of $L$
compared to $p_{r}^{(1)}$,
especially for the preparation and measurement errors.
Thus {\it P}2 has a certain figure of merit when 
the success probability becomes small
($p_s < 0.4$) as seen in Fig. \ref{Threshold}.

As shown above, the present scheme works well even with
a very small success probability $p_s \sim 0.1$,
if the unheralded error probability is $p_{u} \sim 1\times 10^{-4}$.
Since $L$ scales like $1/p_{s}$,
there is a non-zero noise threshold
for a substantially small, but finite success probability.
To compare with Goto's scheme \cite{Goto09}, we set 
$p_{P} = p_{M}=0$ and $p_s=0.9$ $(0.95)$, then
{\it P}1 provides the threshold $p_{u} = 1.6 \times 10^{-3}$ $(2.0 \times 10^{-3})$,
which is comparable or considerably higher than their scheme
especially for a small success probability.
Compared to Dawson's scheme with no photon loss \cite{Dawson06},
the present scheme with $p_s=0.5$ also 
improves the noise threshold.

The memory errors,
which occur on each qubit waiting to be measured,
can be taken into account by adding 
them to the measurement errors.
Since the waiting time is finite,
the memory errors do not deteriorate the performance
crucially provided they are not so large. 
Similarly to the measurement errors, 
the $L$ dependence of the memory errors 
can be reduced by using {\it P}2.

Finally we would like to mention
the resource usage.
For a measure of the total overhead,
we count the average number of PTQGs
consumed per encoded operation on TOWC.
The overhead required 
to construct the star cluster ({\it P}1)
can be described as 
$R_{\rm star} = (L/p_s+L)/p_s^{L}$.
Then, total resources for TOWC can be given by
$R_{\rm tot} =R_{\rm star} R_{\rm TOWC}$, where 
$R_{\rm TOWC}$ indicates the resource consumed per an encoded gate 
on TOWC \cite{Raussendorf07}.
Typically, $R_{\rm TOWC} \sim 10^{7}$ with 
$p_{r} \sim 1\%$ and computation size of $10^{9}$ \cite{Raussendorf07}.
It is worth comparing the present resource usage 
with the conventional schemes \cite{Dawson06,Cho07,Goto09}.
In the case of $p_{s}=0.5$ with $p_{u} = 2 \times 10^{-4}$
and computation size of $10^{9}$,
they amount to 
$R_{\rm star} \sim 7 \times 10^{6}$ and $R_{\rm tot}=R_{\rm star}R_{\rm TOWC} \sim 10^{14}$,
which is significantly (several orders of magnitude)
smaller than the Dawson's and Cho's schemes ($\sim 10^{23}$ and $\sim 10^{18}$, respectively).
In the case of $p_{s}=0.9$ with $p_{u} = 2 \times 10^{-4}$
and encoded gate accuracy of $10^{-4}$,
$R_{\rm star} \sim 30 $ and $R_{\rm tot} \sim 10^{5}$,
which is also considerably
smaller than the Goto's scheme $\sim 10^{7}$.
Simple modifications to the present scheme
would furthermore reduce the resource usage, 
especially for a very small success probability.
Actually, one can improve
the resources for the star cluster 
as $\propto L/p_{s}^{2} (1/p_{s})^{\log L}$
by utilizing the method described in Ref. \cite{Chen06}.
For example, the overhead for the star cluster with $p_{s} = 0.1$ can be 
reduced from $\sim 10^{100}$ to $\sim 10^{11}$.

With an additional overhead,
one would also improve the performance
with respect to the noise threshold.
One can construct a protocol with the independent noise model
by utilizing entanglement purification \cite{DAB03,Goyal06,FY09EP}.
Even with the PTQGs,
entanglement purification works well,
although the resource required for preparing the 
purified star cluster becomes large, but it is still constant overhead.  
Then, one would  completely remove the $L$ dependence in
the renormalized error probability.

{\it Discussion and conclusion.---}
We have investigated fault-tolerant TOWC
with PTQGs.
It has been shown
fault-tolerant computation can be performed well
even with a very small success probability $p_{s} \sim 0.1$,
provided $p_{u} \sim 10^{-4}$.
The present scheme has also succeeded to reduce 
the total overhead considerably. 
The qubit-loss errors, which is another important source of errors
for specific physical systems, such as photonic qubits,
would be treated by following the recent study \cite{Barrett10}.

\begin{acknowledgments}
KF is supported by JSPS Grant No. 20.2157.
\end{acknowledgments}

{\it Note added.---}
During preparation of this manuscript 
we became aware of a related work \cite{Li10},
which also tackles fault-tolerant TOWC
with  typical nondeterministic two-qubit gates
using a different approach for suppressing error accumulation.

\end{document}